\documentstyle[aps,preprint,tighten,epsf,rotate,floats]{revtex}
\newbox\rotbox
\newcommand{\be}{\begin{eqnarray}}
\newcommand{\ee}{\end{eqnarray}}

\def\MeV{\nobreak\,\mbox{MeV}}
\def\GeV{\nobreak\,\mbox{GeV}}

\def\mytoday#1{{}\ifcase\month\or
January\or February\or March\or April\or May\or June\or
July\or August\or September\or October\or November\or December\fi
 \space \number\year}
\begin{document}
\preprint{\vbox{Submitted to {\it Physical Review D }
\hfill ECT*/95-0503 \\
                \null\hfill IFUSP/P-1152}}
\title{Isospin Breaking and Instantons
 in QCD Nucleon Sum Rules}

\vskip 3cm

\author{Hilmar Forkel}
\address{European 
Centre for Theoretical Studies in Nuclear Physics and Related Areas, \\
Villa Tambosi, Strada delle Tabarelle 286, I-38050 Villazzano, Italy} 

\author{and }

\author{Marina Nielsen}
\address{Instituto de F\'\i sica, Universidade de S\~ao Paulo, \\ Caixa 
Postal 66318, 05389--970 - S\~ao Paulo - SP - Brazil }

\date{\today}
\maketitle

\vskip 1 cm
\begin{abstract}

We study isospin breaking instanton corrections to the operator 
product expansion of the nucleon correlation functions. After  
a comparison with quark model calculations based on the 't Hooft 
interaction, we examine the role of instantons in the corresponding 
QCD sum rules. Instanton contributions are found to be absent 
in the chirally even sum rule, but significant in the chirally-odd 
one. They improve the consistency of both sum rules and favor a 
value of the isovector quark condensate close to the chiral 
estimate.

\end{abstract}
\newpage

%%%%%%%%%%%%%%%%%%%%%%%%%%%%%%%%%%%%%%%%%%%%%%%%%%%%%%%%%%%%%%%%%%%%%%%%
%23456789 123456789 123456789 123456789 123456789 123456789 123456789 12

\section{Introduction}

Over the last years growing evidence for a significant role of QCD 
instantons in hadron structure has been collected. It originated 
first from models built on instanton vacuum phenomenology 
\cite{Shu82,DiaPe84} and recently received model independent support 
from cooled lattice studies \cite{ChuHua92}. Indeed, the latter show 
that hadron correlation functions remain almost unchanged if all but 
the instanton fields are filtered out of the equilibrated lattice 
configurations.  

Analytical studies of instanton contributions to the operator product 
expansion (OPE) and to QCD sum rules find a reflection of this picture
in the importance of explicit instanton corrections in the pion 
\cite{for95} and nucleon \cite{for93} channels. The corrections in the 
nucleon channel show a characteristic pattern, which originates from 
the chirality of the quark zero mode states in the instanton background: 
they are small in the chirally-even nucleon correlator and in the 
corresponding sum rule, but significant in the chirally-odd one. 
Indeed, the chirally-odd sum rule could hardly be 
stabilized without instanton corrections, whereas the 
chirally even one is stable and in agreement with phenomenology even 
if the instanton contribution is neglected \cite{Ioffe81}. 

An analogous pattern was found in two recent sum rule calculations of 
the neutron-proton mass difference $\delta M_N $ \cite{yang1,jnp} {\it 
without} instanton corrections, which also show a significant 
discrepancy between the results of the chirally even and 
odd sum rules. Again the former agrees well with phenomenology 
($\delta M_N \simeq 2\MeV$), whereas the latter yields a 
value consistent with zero and thus puts the consistency of the 
two sum rules into question\footnote{Attempts to reduce this 
discrepancy by adding a term attributed to electromagnetic corrections 
to the OPE \cite{yang1} would require a substantial corresponding 
refinement on the phenomenological side of the sum rule, see ref. 
\cite{adami1}.}. This analogy with the nucleon mass sum rules prompted 
us to examine instanton corrections to the isospin violating nucleon 
sum rules, which is the subject of the present paper.

A further, closely related sum rule calculation of isospin violation in 
baryons without instanton corrections \cite{adami1} takes a somewhat 
different approach. The baryon mass splittings are taken as input from 
experiment (after subtraction of the estimated electromagnetic 
contributions), and the two relevant isospin breaking parameters -- 
the quark mass difference $\delta m$ and the difference of up- and 
down-quark condensates $\gamma$ -- are estimated from the sum rules. 
This analysis seems to find consistency between both sum rules, at 
least if the difference $\delta \lambda_N^2$ between the neutron and 
proton pole strengths is fitted, and thus seems incompatible with 
the conclusions of refs. \cite{yang1,jnp}. The fit requires, however, 
an unusually small value of $|\gamma|$, about a quarter of the one 
estimated from chiral perturbation theory, and an uncomfortably large 
continuum contribution. We will come back to this issue below. 

The study of isospin violations in QCD nucleon sum rules can be based 
either on the nucleon correlator in an iospin-violating scalar 
background field \cite{jnp} or on the difference of the neutron and 
proton correlators \cite{yang1,adami1}. We will adopt the latter 
approach. In section \ref{instcorr} we calculate the leading, isospin 
violating instanton corrections to the nucleon correlator, and in section 
\ref{isoviol} we discuss their structure in more detail. Section 
\ref{qmodels} contains a comparison with quark model calculations 
based on instanton-induced interactions. We point out, in particular, 
that the neglect of the vacuum sector in many of these models leads to 
severe limitations in their description of isopin violation effects. 
On the basis of the instanton-corrected nucleon correlators from section 
\ref{instcorr} we then set up the correponding QCD sum rule in section 
\ref{sumrule} and analyze it quantitatively in section \ref{analysis}. 
The final section contains a summary of our results and some conclusions. 

\section{Nucleon correlators}
\label{instcorr}

This section describes the evaluation of small-scale instanton
contributions to the nucleon correlators in the presence of isopin 
breaking. We begin with the correlation function in the proton 
channel, which is characterized by two invariant amplitudes of 
opposite chirality,
\begin{equation}
\Pi_p (q) = i\int d^4{x} \, e^{iq\cdot x}
\langle 0|\, {\rm T} \, \eta_p(x) \, \overline{\eta}_p(0) \,
|0\rangle\ = \rlap{/}{q} \, \Pi_{q,p} (q^2) + \Pi_{1,p} (q^2)\ . 
\label{corr}
\end{equation}

The composite operator $\eta_p$ is built from QCD fields and serves as 
an interpolating field for the proton. Two such (independent) operators 
with minimal mass dimension (i.e. 9/2) can be constructed. We adopt 
the standard choice of Ioffe \cite{Ioffe81}, 
\begin{equation}
\eta_p(x)=\epsilon_{abc}
\left[{u^T_a}(x)C\gamma_\mu u_b(x)\right]
\gamma_5\gamma^\mu d_c(x)\ ,
\label{eta}
\end{equation}
(the neutron current is obtained by interchanging up and down quark 
fields), which allows for a direct comparison with the previous studies
of isospin violation in nucleon sum rules \cite{yang1,jnp,adami1}.

The leading instanton contributions to the correlators can be 
calculated in semiclassical approximation, i.e. by evaluating 
(\ref{corr}) in the background of the instanton field and by taking 
the weighted average of the resulting expression over the quantum 
distribution of the instanton's collective coordinates \cite{for93}. 
These contributions add nonperturbative corrections to the Wilson 
coefficients of the conventional OPE, with which they will be combined 
in section \ref{sumrule}. Isospin breaking originates in this framework 
from the mass difference of up and down quarks,
\begin{equation}
\delta m = m_d - m_u,
\end{equation}
and from the differences in the values of the 
corresponding condensates, 
\begin{equation}
\gamma \equiv \frac{
\langle 0|\overline{d}d-\overline{u}u|0 \rangle}{
\langle 0|\overline{u}u |0 \rangle } \; .
\label{gamma}
\end{equation}

The isovector quark condensate, which determines $\gamma$, is the dominant 
source of nonperturbative isospin violation in the OPE of the correlators, 
since it originates from the lowest dimensional operators with a finite 
vacuum expectation value. The value of $\gamma$ has been estimated in a 
variety of approaches 
\cite{adami1,gasser2,paver1,pascual1,bagan1,dominguez1,dominguez2,narison1}, 
with results varying over almost an order of magnitude, $-1\times 10^{-2} 
\le \gamma \le -2\times 10^{-3} \;$. The sensitivity of the baryon sum 
rule analysis to $\gamma$ can be used for an additional estimate of its 
value \cite{adami1}, which will be adapted to the presence of instanton 
corrections in section \ref{analysis}. For the quark mass difference we 
use the more accurately known standard value $\delta m = 3.3 {\rm MeV}$ 
\cite{gasser2}. 

The rationale behind the semiclassical treatment of instanton 
contributions and the calculational strategy are analogous to those in 
the isosymmetric case \cite{for93}, and we thus just sketch the essential 
steps here. To leading order in the product of quark masses and instanton 
size, instanton effects in the nucleon correlators are associated with 
the quark zero-modes \cite{thooft}
\begin{equation}
\psi_0^\pm(x) = \frac{\rho}{\pi} \frac{1 \pm \gamma_5}{(r^2 + 
\rho^2)^{3/2}} \, \frac{\rlap{/}{r}}{r} \, U, \label{zm}
\end{equation}
where the superscript $\pm$ corresponds to an (anti-) instanton of 
size $\rho$ with center at $x_0$. The spin-color 
matrix $U$ satisfies $(\vec{\sigma} + \vec{\tau}) \, U = 0$ and 
$r = x-x_0$. The zero mode contributions enter the calculation 
of the correlators through the leading term in the spectral 
representation of the quark background field propagator
\begin{equation}
S_{q}^\pm(x,y) = {\psi_0^\pm(x) {\psi_0^\pm}^\dagger(y)\over m_q^*
(\rho)} + O(\rho m_q^*) \; .
\label{zero}
\end{equation}
The flavor dependent effective quark mass $m_q^*(\rho)=m_q-{2\over 3} 
\pi^2\rho^2 \langle\overline{q}q\rangle$ (where $q$ stands for up or
down quarks) in the denominator is generated by interactions with 
long-wavelength QCD vacuum fields \cite{svz2}. Quark propagation 
in the higher-lying continuum modes in the instanton background will 
be approximated as in \cite{for93} by the free quark propagator.

Note that both the zero and continuum mode propagators are flavor 
dependent. The zero mode part contains the effective quark mass, 
which depends on the current quark masses and on the corresponding 
condensates. The current quark masses enter, of course, also the 
continuum mode contributions. 

With the quark background field propagator at hand, the instanton 
contributions to the proton and neutron correlators can now be evaluated. 
As a first, generic result we find that the chirally even amplitudes
$\Pi_q$ for both proton and neutron do not receive leading instanton 
corrections. This generalizes the analogous observation in ref. 
\cite{for93} to finite current quark masses and condensate differences 
and is a consequence of using Ioffe's current. The isopin breaking 
difference of the $\Pi_q$'s for neutron and proton, moreover, vanishes 
for {\it all} interpolating fields, as we will show in the next section. 

The chirally odd amplitudes $\Pi_1$, on the other hand, get sizeable 
instanton contributions, and their difference for proton and neutron 
remains finite. For the proton correlator, to first order in the current 
quark masses and continued to euclidean space-time, we obtain 
\be
\Pi_{1,p}^{ inst} (q^2) = &-& {16\over\pi^4} \int d \rho \, \rho^4 \,
{n(\rho) \,\over{m_0^*}^2(\rho)} \nonumber \\ &\times& \int d^4x \, 
e^{iqx} \left( \frac{1- \zeta}{ N_c} \langle\overline{u}u \rangle
-{i m_u\over\pi^2x^2}\right)\int d^4x_0 {1\over
[(x-x_0)^2+\rho^2]^3[x_0^2+\rho^2]^3} \; ,
\label{inscon}
\ee
where $N_c$ is the number of quark colors. The isoscalar part of the 
effective quark mass in the chiral limit is defined as $m_0^*(\rho)= 
-{2\over 3}\pi^2\rho^2 \langle \overline{q} q\rangle_0$ with $\langle 
\overline{q} q\rangle_0 \equiv (\langle \overline{u} u\rangle+\langle 
\overline{d}d\rangle )/2$ and the dimensionless ratio $\zeta = 
(m_u+m_d)/m_0^*$.

The further evaluation of eq. (\ref{inscon}) requires an explicit
expression for the instanton size distribution $n(\rho)$ in the 
vacuum. Instanton liquid vacuum models \cite{Ver90} and the analysis 
of cooled lattice configurations \cite{ChuHua92} have produced a 
consistent picture of this distribution. The sharply peaked, almost 
gaussian shape of $n(\rho)$ found in ref. \cite{Ver90} can be 
sufficiently well approximated as \cite{shu2}
\begin{equation}
n(\rho) = \bar{n} \,\delta(\rho-\bar{\rho}) \; , \quad \quad \bar{n} 
\simeq \frac12 {\rm fm}^{-4}, \quad \quad \bar{\rho} \simeq \frac13 
{\rm fm}, \label{n}
\end{equation}
which neglects the small half width ($\simeq 0.1 {\rm fm}$) of the 
distribution. In eq. (\ref{n}) we introduced the average instanton 
size $\bar{\rho}$ and the instanton number density $\bar{n}$, which 
equals the density of anti-instantons. $\bar{n}$ can be approximately 
related \cite{cal} to the isoscalar quark condensate by the 
self-consistency condition $\overline{n} = -\frac12 m_0^*(\overline\rho) 
\langle \overline{q} q\rangle_0$, which quite closely reproduces the 
phenomenological value given above and allows the elimination of 
$\overline{n}$ in favor of the quark condensate.

After performing the now trivial integration over instanton sizes, we 
prepare the amplitude (\ref{inscon}) for its use in the corresponding 
sum rule by applying the standard Borel transform \cite{svz}, 
\be
\Pi (M^2) \equiv \lim_{n \rightarrow \infty} \frac{1}{n!} (Q^2)^{n+1} 
\left( - \frac{d}{d Q^2} \right)^n \Pi (Q^2)
\ee
($Q^2 = - q^2$) with the squared Borel mass scale $M^2 = Q^2/n$ kept 
fixed in the limit, and obtain
\begin{equation}
\Pi_{1,p}^{ inst} (M^2) = - {3\over 4\pi^2}\left[ \frac{1- \zeta}{N_c} 
\langle\overline{u}u\rangle M^4 I_1(M^2 \bar{\rho}^2) - \frac{1}{16\pi^2}
m_u \bar{\rho}^4 M^{10} I_2(M^2 \bar{\rho}^2) \right] \; , \label{incorr}
\end{equation}
in terms of the two dimensionless integrals
\begin{equation}
I_1(z^2) = \int_{z^2/4}^\infty dx \, {x^2\over(x-z^2/4)^2}
e^{-x^2 (x-z^2/4)^{-1}} \ , \label{int1}
\end{equation}
\begin{equation}
I_2(z^2) = \int_0^\infty dx_1 \int_0^{z^{-2}} dx_2 \,
{x_2^2(z^{-2}-x_2)^2\over (x_1+x_2-z^2x_2^2)^5}
e^{-{1\over4} (x_1+x_2-z^2x_2^2)^{-1} } \; . \label{int2}
\end{equation}
The amplitude for the neutron follows from (\ref{incorr}) by 
interchanging up and down quark masses and condensates. Eq. 
(\ref{incorr}) generalizes the isospin-symmetric amplitude of ref. 
\cite{for93}, which is recovered in the chiral limit.

The two integrals (\ref{int1}) and (\ref{int2}) contain the instanton 
corrections to the Borel transformed Wilson coefficients of the unit 
operator (in $I_2$) and of $\langle\overline{u}u\rangle$ (in $I_1$). 
Additional contributions from quark modes with momenta below the 
renormalization scale $\mu$ of the OPE should be excluded from 
$I_1$ and $I_2$ in order to avoid double counting of the physics 
contained in the condensates. Fortunately, the instanton background 
field induces only one soft contribution up to operators 
of dimension 6, corresponding to a four-quark condensate in $I_2$. We 
will correct for this contribution in section \ref{sumrule}.

\section{Isopin violation and instantons}
\label{isoviol}

It is instructive to analyze the isospin properties and the origin of 
isospin breaking in the instanton induced amplitude (\ref{incorr}) in 
more detail. This analysis will also lay the foundation for our 
discussion in the next section, where we clarify some crucial 
differences between our approach and quark model calculations. These 
differences explain, in particular, the absence of instanton induced 
contributions to $\delta M_N$ in many quark models, contrary to our 
findings in the correlator approach. 

The gluonic sector and the quark-gluon vertex of QCD are both flavor 
independent. The structure of the interaction with the instanton 
background field is thus the same for up and down quarks and the 
background field propagator (\ref{zero}) is diagonal in isopin space. 
Its only flavor dependence enters through the current quark masses and 
condensates in $m_q^*$. This has characteristic consequences for the 
instanton-induced interactions between quarks, which generate the 
correlator amplitude (\ref{incorr}) and can be extracted from the 
calculation in section \ref{instcorr}: 
\be
\langle 0 | &T& q_{A, \alpha, a}  (x_1)\, \bar{q}_{B, \beta, b} (y_1) \, 
q_{C, \gamma, c} (x_2) \, \bar{q}_{D, \delta, d} (y_2) \, | 0 \rangle 
\nonumber \\ &=&  (\delta_{A B} \delta_{C D} - \delta_{A D} \delta_{C B}) 
\int d \rho \frac{n(\rho)}{(m^*_u \rho) (m^*_d \rho)} (2 \pi^2 \rho^3)^2 
\nonumber \\ & & \times \int d^4 x_0 \, C(r_1) C(u_1) C(r_2) C(u_2) 
\sum_{L/R} << (P_{L/R} \tilde{P}_{a b})_{\alpha \beta} (P_{L/R} 
\tilde{P}_{c d})_{\gamma \delta} >>_{{\rm SU(3)}_c}. \label{ivertex}
\ee

In this expression, capital latin, greek and small latin indices refer 
to isospin, Dirac spin and color, respectively. The brackets $<<...>>$ 
indicate the average of the instanton's color orientation over the Haar 
measure of ${\rm SU(3)}_c$. The chiral projection operators are $P_{L/R} 
= (1 \pm \gamma_5)/2$, the distances from the instanton center are 
denoted $r_i = \sqrt{(x_i-x_0)^2}$ and $u_i = \sqrt{(y_i-x_0)^2}$, the 
nonlocality of the vertex is contained in the functions 
\be
C(r) = \left( \frac{r^2}{r^2 + \rho^2} \right)^{\frac32} ,
\ee
and its color structure is given by the spin-color tensor
\be
\tilde{P}_{\alpha \beta,a b} = \delta_{\alpha \beta} 
\left( \begin{array}{cc} 1 & 0 \\ 0  & 0 \end{array} \right)_{a b}
- \,\, \vec{\Sigma}_{\alpha \beta} \left( \begin{array}{cc} 
\vec{\tau} & 0 \\ 0  & 0 \end{array} \right)_{a b} ,
\ee
which makes the embedding into an SU(2) subgroup of ${\rm SU(3)}_c$
explicit and contains the characteristic spin-color coupling. Lorentz 
and color covariance become manifest only after averaging over the 
color group. 

The nonlocal four-quark vertex (\ref{ivertex}) originates from the 
quark zero modes in the 
instanton field and was first derived by 't Hooft \cite{thooft} for 
$m_q^* = m_q$. It has, separately for both quark chiralities, a
well-known determinantal flavor structure and is thus $SU(2)_L 
\times SU(2)_R$ and in particular isospin symmetric\footnote{In 
addition it breaks the axial $U_A(1)$ symmetry, which is a reflection 
of the Adler-Bell-Jackiw anomaly and a celebrated instanton effect 
\cite{thooft}.}. Even if the flavor dependent quark masses and 
condensates enter the vertex explicitly, isospin violation can thus
not originate solely from the instanton induced interaction. In 
the OPE, however, this vertex generates nonperturbative contributions 
to the Wilson coefficients, which can either become themselves isopin 
violating due to the finite quark mass difference or multiply isopin 
violating operators. Examples for both of these cases were found in 
section \ref{instcorr}.

At short distances, multi-instanton contributions to the nucleon 
correlators are suppressed since $x/\bar{R} \ll 1$. Additional 
corrections, which originate from only one valence quark propagating in 
a zero mode state and generate contributions to the quark self-energy, 
are subleading in $m^* \bar{\rho}$ and not considered in this paper. 

The structure of the instanton-induced four-point function 
(\ref{ivertex}) explains some qualitative features of the correlator
(\ref{incorr}). The definite chirality of the quark legs, inherited from 
the zero mode states, lets this vertex act only between quarks 
which are coupled to spin 0 in the interpolating fields, and the 
Dirac structure of the instanton contribution to the nucleon correlator 
is thus determined by the remaining ``valence'' quark line. Only the 
contribution to the Wilson coefficient of the unit operator, in which 
this line is in a non-zero-mode state (approximated by the free quark 
propagator), can thus generate a chirally even amplitude. 
According to our discussion above, however, such contributions are 
isospin conserving and this explains why the difference of the 
chirally-even amplitudes of the neutron and proton correlators is not 
corrected by instantons for any choice of the interpolating field. The 
instanton contributions to the chirally even amplitudes of neutron 
and proton vanish individually only for the Ioffe current, as already
pointed out.

Another characteristic feature of the correlator (\ref{incorr})
follows directly from the flavor structure of the 't Hooft vertex.
Since only one quark pair of each flavor can take part in the zero mode 
induced interaction (\ref{ivertex}), the valence quark in the 
non-zero-mode state in the proton (neutron) correlator must be an up 
(down) quark. This explains why we find only contributions proportional 
to the up quark mass and condensate in eq. (\ref{incorr}). 

To summarize, instanton induced interactions contribute to isopin breaking 
in the operator product expansion of the nucleon correlators (in the 
framework of our approximations) in two distinct ways: they correct the 
Wilson coefficient of the unit operator, which becomes isopin dependent 
due to the difference of the current quark masses, and they contribute 
to the coefficients of the isospin violating operators $\bar{u}u$ and 
$\bar{d}d$. Both of these corrections affect only the chirally odd 
amplitude of the correlators. 

\section{Comparison with quark models}
\label{qmodels}

Instanton induced interactions have been included in several quark model 
calculations of mass splittings in baryon iso-multiplets. It is useful 
to compare the results of such calculations to those of our approach. 
To be specific, we will base this comparison on studies in the MIT bag 
model \cite{Yan84,Dor88}, which deals with relativistic, light quarks 
and is in this respect similar to the correlator approach. Most of our 
conclusions, though, will apply to a wider range of quark models. 

Bag model calculations of hadron mass shifts due to instantons 
\cite{Hor78} are based on a localized version of the 't Hooft 
interaction, cast into the form of an effective 
lagrangian. Indeed, the pointlike limit of the vertex (\ref{ivertex}) 
(in Minkowski space) is reproduced by the lagrangian \cite{svz2}
\be
{\cal L}_{inst} &=& - (\frac43 \pi^2 \bar{\rho}^3)^2 
\frac{\bar{n}}{(m^*_u \bar{\rho}) (m^*_d \bar{\rho}) }  \sum_{L/R}
\Bigg\{ (\bar{u}_R u_L) (\bar{d}_R d_L)  \nonumber \\  &+&  
\frac{3}{32}  \left[ (\bar{u}_R \lambda_a u_L) (\bar{d}_R \lambda_a 
d_L) - \frac34 (\bar{u}_R \lambda_a \sigma_{\mu \nu} u_L) (\bar{d}_R 
\lambda_a \sigma^{\mu \nu} d_L) \right] \Bigg\}, \label{efflagr}
\ee
which is obtained from (\ref{ivertex}) by amputating the external 
quark propagators, neglecting the nonlocality due to the finite 
instanton size, performing the average over the color orientation 
of the instanton\footnote{As long as this lagrangian is evaluated 
only in color singlet states, one could of course skip the color 
averaging and use the neither Lorentz nor ${\rm SU(3)}_c$ invariant 
version instead, with identical results.}, specifying the instanton 
density in the form (\ref{n}) and continuing back to Minkowski 
space-time. 

The shrinking of the instanton vertex to its pointlike limit will 
probably not cause significant errors in bag model results for 
low-lying hadrons. This is because quarks in the bag can separate up to 
the diameter $2R \sim 2 \,{\rm fm}$, so that their inverse momenta in the 
ground state are considerably larger than the average instanton size
which characterizes the extent of the vertex. Note, however, that the 
nucleon correlator in QCD sum rule calculations is probed at an order 
of magnitude smaller distances, where the details of the short distance 
dynamics and thus the nonlocality of the vertex become important.

Besides the structure of the instanton induced quark interaction,  
bag calculations share some other common features with the nucleon 
correlators (\ref{corr}), notably in the construction of the nucleon 
states. The spin, color and flavor structure of the bag model (i.e. 
SU(6)) proton state,
\begin{equation}    
| p, \uparrow \rangle = \frac{1}{\sqrt{18}} \epsilon_{a b c } \left[ 
( u^+_{a \downarrow} d^+_{b \uparrow} -  u^+_{a \uparrow} d^+_{b 
\downarrow} ) u^+_{c \uparrow} \right] |0 \rangle \label{su6}
\end{equation}
(Arrows indicate the value of the total spin projection $j_3$ of the 
quarks (in the bag ground state) and of the proton.) and the 
corresponding one for the 
neutron (which is obtained from $- | p \rangle$ by interchanging up 
and down quarks) is essentially identical to that of the interpolators 
(\ref{eta}). This is of course just a consequence of the fact that both 
are constructed to carry nucleon quantum numbers, which ensures that 
they have identical properties under Lorentz, color, isospin and the 
standard discrete transformations. 

Despite these similarities, bag model calculations do not find any 
instanton contribution to the proton-neutron mass 
difference\footnote{as long as one-zero-mode corrections (see above) 
are neglected} \cite{Yan84,Dor88}. The neutron and proton 
mass shifts induced by (\ref{efflagr}) are evaluated in first order 
perturbation theory between the SU(6) states. A straightforward
calculation gives
\be
\langle p | &-& \int d^3 x \, {\cal L}_{inst}(x) \,\, | p \rangle \, = 
\, (\frac43 \pi^2 \bar{\rho}^3)^2 \, \frac{\bar{n}}{(m^*_u \bar{\rho}) 
(m^*_d \bar{\rho}) } \, \left(\frac{N_u N_d}{8 \pi} \right)^2 \nonumber 
\\ 
&\times& \int d^3 x \left\{ 5\, (a^2_1 - a_2^2)^{(u)} (a^2_1 - 
a_2^2)^{(d)} + 4\, (a^2_1 + a_2^2)^{(u)} (a^2_1 + a_2^2)^{(d)} + 
20\, a_1^{(u)} a_2^{(u)} a_1^{(d)} a_2^{(d)} \right\}, \label{bagme}
\ee
(the integrations extend over the bag volume) where the ground state 
quark wavefunction in the bag is written as
\be
\psi^{(q)}_{j_3} = \frac{N_q}{\sqrt{4 \pi}} \left(\begin{array}{c} 
a_1^{(q)} (r)  \\ i \vec{\sigma} \vec{r}\, a_2^{(q)} (r) \end{array} 
\right) \chi_{j_3}
\ee
with $a_1^{(q)} = \sqrt{\frac{\omega_{q}+m_q}{\omega_{q}}} \, j_0 
\left(\frac{\kappa_{q} r}{R}\right)$ and $a_2^{(q)} = 
\sqrt{\frac{\omega_{q}-m_q}{\omega_{q}}} \,j_1 \left(\frac{\kappa_{q} 
r}{R} \right)$. ($R$ is the bag radius, $N_q$ a normalization constant, 
$\chi_{j_3}$ are the Pauli spinors and $\omega_{q}$, $\kappa_{q}$ are 
the energy and (dimensionless) momentum quantum numbers of the quark 
ground state.) The result for the neutron matrix element can be 
immediately infered from (\ref{bagme}) by exchanging up and down quark 
operators in the proton states. Since ${\cal L}_{inst}$ is symmetric 
under $u \leftrightarrow d$, one can as well exchange $u$ and $d$ 
everywhere in the expression for the proton matrix element. Eq. 
(\ref{bagme}), however, is manifestly invariant under this exchange  
and therefore the difference of the matrix elements indeed vanishes:
\begin{equation}
\delta M^{inst}_{N,bag} = \langle p | \int d^3 x \,
{\cal L}_{inst} \, | p \rangle - \langle n | \int d^3 x \,  
{\cal L}_{inst} \, | n \rangle  = 0. \label{bagdif}
\end{equation}

This is in contrast to the result for the Borel transformed nucleon 
correlators, 
\be
\Pi_{1,n}^{ inst} (M^2) - \Pi_{1,p}^{inst} (M^2) \neq 0, \label{diff}
\ee
which can be translated via a QCD sum rule into a finite instanton 
contribution to the mass difference (see below).

Of course, the bag model contains ad hoc assumptions on quark 
confinement, breaks chiral symmetry explicitly and differs also in other 
aspects from the model-independent correlator approach. 
One would thus not expect the results to agree quantitatively. It is 
at first surprising, however, to find a qualitative difference, namely 
the exact absence of {\it any} instanton mediated mass shift $\delta 
M^{inst}_{N,bag}$. In view of the similarity of the interactions 
(\ref{ivertex}) and (\ref{efflagr}) one is lead to search for the 
origin of this difference in the description of the nucleon states. 
And indeed, here the two approaches differ crucially.  

A first difference is that the quarks in the bag matrix elements are 
restricted to their ground states with total spin $j = 1/2$ (i.e. to 
$l = 0$ or $1$), whereas the interpolators can create quarks in all 
orbital angular momentum states. In fact, the created pointlike wavepacket 
has overlap with the whole tower of excited states carrying nucleon quantum 
numbers, including states in the many-particle continuum. Experience 
from QCD sum rules \cite{Ioffe81} and from lattice data 
\cite{Chu93,Lei94} shows, however, that already at rather small 
distances a main contribution to the nucleon correlator comes from 
the nucleon ground state. This tendency is further enhanced
by the Borel transform, which exponentially suppresses contributions 
from higher lying states. One does therefore not expect these states 
to contribute significantly to a finite value of the difference 
(\ref{diff}) in the fiducial Borel mass domain (see below), let alone 
to be its only cause.

Indeed, the crucial difference between the bag and correlator results 
can rather be traced to the description of the nucleon ground states 
themselves, and in particular to their flavor content. While the SU(6) 
states (\ref{su6}) (as well as the interpolating fields (\ref{eta})) 
have good isospin, this is {\it not} the case for the states $\eta_{p,n} 
|0 \rangle$ which are created by the interpolators and studied in the 
correlator (\ref{corr}). 

Virtual ``sea'' quarks and other perturbative and nonperturbative 
vacuum fluctuations give these states a much richer (and more realistic) 
flavor structure. They inherit, in particular, isospin breaking components 
from the vacuum fields. The short distance part of this nontrivial flavor
content is captured both in the OPE and in the instanton corrections and 
causes (\ref{diff}) to be finite. At very short distances it originates 
from the quark mass differences, and at larger distances it enters 
predominantly through the difference between up and down quark 
condensates. The neglect of both of these ingredients in the isospin 
structure of the bag wavefunctions leads, on the other hand, to the 
symmetry of the proton matrix element (\ref{bagme}) under exchange of 
the two quark flavors in the states (\ref{su6}) and thus to the 
absence of an instanton-induced mass difference, eq. (\ref{bagdif}). 

In ref. \cite{Dor88} an attempt was made to include long-wavelength 
vacuum fields into the bag interior. Long and short distance physics 
inside the bag, however, cannot be reliably separated. In particular,
such a scale separation (which is indispensable to control the 
interactions with the vacuum fields) cannot be based on a short distance 
expansion, which would in fact badly diverge at distances of the order 
of the bag radius $R \sim \Lambda_{QCD}^{-1}$. Higher order interactions 
with 
the background fields (leading to contributions from higher dimensional 
condensates) are thus not suppressed, and it is hard to see how their
neglect can be justified and how double counting of quark physics 
can be avoided. In addition, both the contributions of the interactions 
with the long-wavelength background fields and with the instanton to the 
matrix elements are calculated independently to leading order and then 
added. Combined effects of instantons and the other vacuum fields, as 
described by the OPE, are therefore still lacking\footnote{except for 
the factor $(m^*_u m^*_d)^{-1}$ in the instanton-induced interaction 
(\ref{efflagr})} and eq. (\ref{bagdif}) remains to hold.

The large distance scales over which quarks can interact are a generic 
bag model problem, since clearly not all nonperturbative physics can be 
absorbed into the boundary conditions. For QCD sum rule calculations, 
on the other hand, a reliable description of the correlation functions 
up to distances of about $0.2 \,{\rm fm}$ is usually sufficient. At 
these rather small distances the long-wavelength physics can still be 
controlled in a model independent way by only a few generic and physical 
parameters, the low-dimensional condensates. 

Instanton physics supplies, in fact, yet another example for the problems 
with treating quark interactions in the large bag interior. The 
neglect of multi-instanton effects (beyond the mean field level) is a 
basic assumption underlying the interactions (\ref{ivertex}) 
and (\ref{efflagr}). This approximation can be justified in the correlator 
for distances $x \ll \bar{R}$, but hardly at the scales set by the bag 
diameter. Instanton liquid simulations \cite{Schae93} indeed confirm that
multi--instanton effects become important at distances of the order of 
the average separation between instantons, $x \ge \bar{R} \simeq 1\, 
{\rm fm}$.

To conclude, quark and bag model calculations which restrict the 
evaluation of instanton-induced baryon mass shifts to calculating 
the expectation value of the effective 't Hooft lagrangian between
SU(6) eigenstates miss important sources of isospin asymmetry, 
in particular from isospin violating vacuum fields. This physics, which
significantly affects the estimates of mass splittings in baryon 
iso-multiplets, is however captured in the instanton-corrected OPE of 
the nucleon correlators. 

\section{Isopin violating nucleon sum rules}
\label{sumrule}

In this section we combine the instanton contributions to the nucleon 
correlation functions with the conventional operator product expansion 
\cite{yang1} and set up the QCD sum rule for the difference of the 
neutron and proton correlators. 
 
Since the average instanton size $\bar{\rho}$ is smaller than the 
inverse renormalization point $\mu^{-1} \simeq 0.4 \, {\rm fm}$ of 
the OPE, the major part of the instanton corrections will contribute 
to the Wilson coefficients. These corrections can be directly added 
to the standard OPE, since they originate from nonperturbative physics 
which was not previously accounted for. 

The integral $I_2$ in eq. (\ref{incorr}), however, contains besides 
the instanton contribution to the Wilson coefficient of the unit 
operator also a soft part, as pointed out in section \ref{instcorr}. 
It originates from the region in loop momentum space where the hard 
external momentum $Q^2$ is carried exclusively by the quark line which 
is not participating in the zero-mode induced interaction. No such 
contribution is contained in $I_1$, since here the third quark line 
interacts with the quark condensate and thus does not carry momentum.

The soft part of $I_2$ represents an instanton contribution to the 
four-quark condensate. Fortunately this is the only operator up to 
dimension 6 which receives such a correction. Indeed, a general 
theorem \cite{Dub81} severely limits the number of condensates in 
the OPE of hadronic correlators which can be induced by self-dual 
background fields. 

In order to prevent double counting of long-wavelength physics, the 
four-quark condensate terms in the OPE and the instanton contributions 
have to be adapted before combining them with the other OPE terms of
$\Pi_1$. As in ref. \cite{for93} we will neglect the comparatively 
small OPE contribution to $\langle\overline{u}u \overline{d}d\rangle$ 
and keep instead the contribution induced by the 't Hooft vertex in 
the limit of vanishing external momenta. A more accurate procedure, 
namely to subtract explicitly the part of the instanton contribution
which originates from momenta below the renormalization scale, will be 
described elsewhere \cite{for952}. Four-quark condensates of the type  
$\langle\overline{u}u\overline{u}u\rangle$ and 
$\langle\overline{d}d\overline{d}d\rangle$, on the other hand, do not 
receive single--instanton contributions and remain unchanged.

After implementing the above modification, we can combine the instanton 
part (\ref{incorr}) with the OPE of the chirally odd sum rule of 
ref.~\cite{yang1}. Taking the difference of the neutron and proton 
sum rules and transfering the continuum contributions to the left-hand 
(i.e. OPE) side, we obtain
\begin{eqnarray}
{M^6\delta m\over 16\pi^4}E_2L^{-8/9}-{\gamma\over
4\pi^2}\langle\overline{q}q\rangle_0 M^4E_1 + {4
\over 3}\delta m\langle\overline{q}q\rangle_0^2
&+& {\gamma\over4\pi^2}\langle\overline{q}q\rangle
M^4I_1(z^2)-{3\over 64\pi^4}\delta m
\overline{\rho}^4M^{10}I_2(z^2)L^{-8/9}
\nonumber\\*[7.2pt]
=\left[ 2\lambda_N^2{M_N^2\over M^2} \delta M_N -
\lambda_N^2\delta M_N - \delta\lambda_N^2 M_N 
 \right]&e&^{-M_N^2/M^2} 
-{1\over4\pi^2}\langle\overline{q}q\rangle_0 s_{1} e^{-s_{1}/M^2}
\delta s_{1}\; ,
\label{sum_1}
\end{eqnarray}
where $M_N=(M_p+M_n)/2$ and $\lambda_N=(\lambda_p+\lambda_n)/2$
denote the isoscalar nucleon mass and coupling to the interpolating
field. (We neglect the small gluon condensate contribution.) 
The isospin violating differences of the overlap and threshold 
parameters are $\delta\lambda_N^2 =\lambda_n^2-\lambda_p^2$,
$\delta s_{1}=s_{1n}-s_{1p}$, and the factor $L^{-8/9}$, with
 $L=\ln(M^2/\Lambda_{\rm QCD}^2)/\ln(\mu^2/ \Lambda_{\rm QCD}^2)$ 
and $\Lambda_{\rm QCD}=150\MeV$, accounts for the anomalous 
dimensions of the composite operators and sets their 
renormalization point to $\mu=500\MeV$. 

The contributions from the continuum, starting at the effective
threshold $s_1$, are as usual combined with the leading OPE term and 
described by the functions $E_1\equiv 1-e^{-s_1/M^2}\left({s_1\over
M^2}+1\right)$ and $E_2\equiv 1-e^{-s_1/M^2}\left({s_1^2\over 2M^4}
+{s_1\over M^2}+1\right)$ \cite{Ioffe81}. Their definitions are 
identical to those in the individual sum rules. Note that additional 
terms in the difference sum rules, proportional to $\delta s_1$ and 
$\delta s_q$ (see below), originate from the continuum terms of the 
individual sum rules. They do not correspond to the cut structure 
(i.e. to the leading OPE behaviour) of the difference sum rule, 
however, and are thus not needed to match the large-$s$ behavior of 
the OPE. 

With the standard definitions 
$a\equiv-4\pi^2\langle\overline{q}q\rangle_0$, and $\tilde{\lambda}_N^2 
\equiv 32\pi^4\lambda_N^2$, our sum rule eq. (\ref{sum_1}) now assumes 
its final form
\begin{eqnarray}
 e^{M_N^2/M^2}\left[M^8\delta mE_2L^{-8/9}\right.&+&\left.
M^6\gamma a E_1 +{4\over 3}\delta m M^2 a^2  
- M^6\gamma a I_1(z^2) - {3\over 4}\delta m\overline{\rho}^4M^{12}
I_2(z^2)L^{-8/9}\right] \nonumber\\*[7.2pt]
=\tilde\lambda_N^2 M_N^2 \delta M_N &-&\left({\tilde\lambda_N^2 
\over 2}\delta M_N + {\delta\tilde\lambda_N^2\over2}M_N\right)M^2  
+ a s_1 M^2 e^{-(s_{1}-M_N^2)/M^2}\delta s_1  \; .
\label{fi1}
\end{eqnarray}

The corresponding sum rule from the $\rlap{/}{q}$ structure 
\cite{yang1,jnp,adami1} is unaffected by leading instanton 
corrections, 
\begin{eqnarray}
 e^{M_N^2/M^2}\left[-aM^4\delta m E_0 L^{-4/9}\right.&-&\left.
{4\over 3}M^2\gamma a^2 L^{4/9} + {m_0^2\over 6}\delta m M^2
a L^{-8/9} + {m_0^2\over 3}\gamma a^2
L^{-2/27}\right] \nonumber\\*[7.2pt]
=\tilde\lambda_N^2 M_N \delta M_N &-&
{\delta\tilde\lambda_N^2\over2}M^2 +{1\over 4}\left(s_q^2+
{b\over2}\right) M^2 e^{-(s_{q}-M_N^2)/M^2}L^{-4/9}\delta s_q  \; .
\label{fiq}
\end{eqnarray}
The two parameters $b=\langle g_s^2 G_{\mu\nu}^a G^{\mu\nu}_a\rangle=
0.5\GeV^4$ and $m_0^2\equiv\langle g_s\overline{q}\sigma\cdot G 
q\rangle/ \langle\overline{q}q\rangle_0 = 0.8\GeV^2$ are fixed at 
their standard values and $E_0\equiv 1-e^{-s_q/M^2}$. 

Above we have written both sum rules, eqs. (\ref{fi1}) and (\ref{fiq}), 
in their most general form, which allows for independent values of 
the isosymmetric continuum thresholds $s_1$ and $s_q$. Below we will, 
however, follow the standard practice and set $s_1=s_q \equiv s_0$.

At this point it might be useful to recall the main assumptions and 
approximations which went into the OPE of these sum rules and into the 
parametrization of their phenomenological sides. (For more details see 
\cite{svz2,adami1}.) The short-distance expansion is carried out up to 
operators of dimension six and the perturbative part of the Wilson 
coefficients is calculated to leading order in the strong coupling 
$\alpha_s$. Less systematic uncertainties arise from the not precisely 
known values of the condensates and from the standard factorization of 
the four-quark condensates. 

As is common practice in sum rule calculations, the Wilson coefficients 
are calculated without explicit infrared cutoff since at scales up to 
about $0.5 - 1 \,{\rm GeV}$ non-perturbative contributions 
strongly dominate over the perturbative ones. The explicit removal of the 
latter becomes therefore practically unnecessary \cite{nov85}. For 
the same reason, the condensates (and the quark masses) depend in the 
above range rather weakly on the renormalization scale. Even without 
explicitly specifying the infrared regularization scheme of the Wilson 
coefficients scale--dependent quantities are understood to be taken at 
a $\mu$ in the above range, and we use specifically $\mu=0.5\, {\rm GeV}$.

On the phenomenological side the main assumption is that of local duality 
between the hadron and quark--gluon descriptions of the continuum. It 
has been found to work well in many sum rule studies and also in recent 
lattice simulations of point-to-point correlators \cite{Chu93}. In our 
context it is put to a harder test since we consider differences of two 
spectral functions. Here even more than in the single nucleon sum rules 
the exponential Borel suppression of continuum contributions is important 
in order to increase the sensitivity of the sum rules to the ground state 
contributions.

In the numerical evaluation of the sum rules the upper limit of the 
Borel interval is determined such that contributions from the continuum 
do not exceed a given percentage of the full OPE contribution. Otherwise 
the sum rules would be relatively less sensitive to the pole contribution 
of interest and a good fit quality would become a trivial consequence of 
continuum domination (instead of being a consistency criterion), since 
the continuum is modeled after the leading OPE behavior. Moderate 
continuum contributions are therefore a necessary condition for reliable 
sum rules, and in the next section we will check how these contributions 
are affected by the instanton terms. 

\section{Quantitative sum rule analysis}
\label{analysis}

The quantitative analysis of isospin violation in the nucleon sum 
rules aims at determining the isospin breaking parameters on the 
phenomenological side from the best fit to the ``theoretical'' 
left hand side. Taking all the other parameters from 
the standard, isosymmetric nucleon sum rules or from experiment, 
it would still require a four-parameter fit to determine $\delta M_N, 
\delta\tilde\lambda_N^2, \delta s_q$ and $\delta s_1$ independently. 
Limitations in the parametrization of the spectral densities and 
approximations on the theoretical side would, however, make such a 
fit unstable both with and without instantons. 

In order to reduce the number of fit parameters, one is thus led to 
either fix the only phenomenologically known one, $\delta M_N$, at its 
experimental value \cite{adami1} or to make assumptions relating at 
least two of the remaining isospin breaking parameters. The authors 
of ref. \cite{yang1}, for example, assume $\delta s_q = \delta s_1$ in 
their analysis or, alternatively, neglect differences in the effective 
continuum of proton and neutron channels entirely, {\it i.e.}  $\delta 
s_q = \delta s_1 = 0$. Since such assumptions lack theoretical 
foundation, the associated errors can not be reliably estimated or 
controlled. We thus prefer to follow the approach of ref. \cite{adami1}, 
taking $\delta M_N^{non-elm} = 2.05 \pm 0.30 \MeV$ as input from 
phenomenology. This value is derived from the experimental 
mass difference $\delta M_N^{ept} = 1.29\MeV$ \cite{data} by subtracting 
the electromagnetic contribution $\delta M_N^{elm} = -0.76 \pm 0.30\MeV$  
\cite{gasser3}.

For the isoscalar nucleon mass and quark condensate we use the standard 
values  $M_N=940\MeV$ and $\langle\overline{q}q\rangle_0 = -(225\MeV)^3$. 
The residuum of the isosymmetric nucleon pole, $\tilde{\lambda}_N^2 = 1.8 
{\rm \GeV}^6$, and the isospin average of the continuum threshold, $s_0= 
{\rm 2.2\GeV^2}$, are obtained from the instanton-corrected nucleon 
mass sum rules \cite{for93} in the same Borel window as the one used 
below.

The isospin-breaking parameters $\delta\tilde\lambda_N^2$, $\delta 
s_1$ and $\delta s_q$ are then calculated by minimizing the difference 
between the left- and right-hand sides of the sum rules (\ref{fi1}) 
and (\ref{fiq}) under the logarithmic measure $\delta$ of ref. 
~\cite{Ioffe81} in the  Borel-mass region $0.8\GeV^2 \le M^2 \le 
1.4\GeV^2$.

We performed this minimization for various values of $\gamma$ in the 
range $-1\times 10^{-2} \le \gamma \le -2\times 10^{-3} \;$ discussed 
in section \ref{instcorr}. We find a better agreement between both sum 
rules towards larger (and more conventional) values of $|\gamma|$ in 
this interval, whereas the analogous study of \cite{adami1}, 
which neglects instanton contributions, prefers an unusually small 
value,  $|\gamma| = 2 \times 10^{-3}$. For $\gamma = - 1 \times 
10^{-2}$, in particular, our best fit between LHS and RHS results in 
the parameter values
\begin{equation}
\delta \tilde\lambda_N^2 = -2.1\times 10^{-4}\GeV^6, \; \; \; 
\delta s_1 = -1.7\times 10^{-2}\GeV^2,  \; \; \; 
\delta s_q = 1.03\times 10^{-3}\GeV^2 .
\end{equation}

The Figures 1 - 3 show different aspects of this fit. In 
order to compare the fits of the optimized $\Pi_q$ and $\Pi_1$ sum 
rules, we transfer all but the first two terms on the RHS of the
$\Pi_q$ sum rule (\ref{fiq}) to the left and we rewrite the $\Pi_1$ 
sum rule (\ref{fi1}) analogously, so that the same two terms, 
$\tilde\lambda_N^2 M_N \delta M_N -{\delta\tilde\lambda_N^2\over2}M^2$,
remain on its RHS. Figure 1 compares this RHS (continuous line) with 
the modified LHS of the $\Pi_q$ (dotted line) and $\Pi_1$  (dashed 
line) sum rules.

In Figure 2 we plot the resulting neutron-proton mass difference 
$\delta M_N$ as a function of $M^2$, obtained by solving both  
optimized sum rules for $\delta M_N(M^2)$. These curves show an 
extended stability plateau, which confirms the satisfactory 
agreement between the two sum rules. Indeed, this Borel mass 
independence of observables is the only intrinsic consistency 
criterion for the sum rules.

In order to compare the relative size and behavior of the 
OPE and instanton contributions to the $\Pi_1$ sum rule (\ref{fi1}) 
(recall that the $\Pi_q$ sum rule does not receive instanton 
corrections), we display both of them separately, as well as their 
sum and the fit to the optimized RHS of eq. (\ref{fi1}), in Figure 3. 
The instanton terms reach almost the magnitude of the perturbative 
and power terms and play clearly an important role in determining 
the sum rule results. The usual practice to neglect these 
contributions seems thus unjustified.

It is also instructive to compare our results of Figures 1 and 2 with 
the analogous curves, but calculated without instanton corrections. 
Recall that in this case the $\langle \overline{u} 
u \overline{d} d\rangle$ part of the four-quark condensate has to be 
restored in eq. (\ref{fi1}), which changes the factor $4/3$ in its 
Wilson coefficient to $-2/3$. As already noted a smaller absolute 
value of $\gamma$ is favored in this case, and the curves in Figures 4 
and 5 were obtained by optimizing the sum rules with $\gamma = - 2 
\times 10^{-3}$. Up to small corrections from the neglected 8-dimensional 
condensates, they correspond to the ones\footnote{The sum rule of ref. 
\cite{adami1} contains an error in the coefficient of the four-quark 
condensate which we have corrected.} analyzed in \cite{adami1}. 

From Figure 4 it is also clear that rather different values of the isospin 
breaking parameters ($\delta \tilde\lambda_N^2 = 1.4 \times 10^{-2} 
{\rm GeV}^6, \delta s_q = 7.0 \times 10^{-3} {\rm GeV}^2, \delta s_1 
= 1.2 \times 10^{-2} {\rm GeV}^2$) are required to fit phenomenological 
and theoretical sides as long as instanton contributions are neglected. 
The difference between the pole strength of neutron and proton, in 
particular, becomes about two orders of magnitude larger and changes
sign. 

More importantly, however, the small modulus of $\gamma$ preferred by 
this fit has an unwelcome consequence. Closer inspection of the sum 
rules reveals that decreasing values of $| \gamma |$ lead to increasing 
contributions from the continuum relative to the power corrections in 
the optimized sum rules. Indeed, the parameter values used above 
correspond to a continuum contribution of 90\%  in the chirally-odd 
sum rule (and about 37\% in the chirally-even sum rule). This continuum 
domination casts serious doubts on the reliability of the 
chirally-odd sum rule, even if fit quality and stability seem 
satisfactory ({\it cf.} Fig. 5). In both instanton-corrected sum rules, 
on the other hand, the continuum contributions remain moderate (about 
20\%). 

It is interesting to note that another recent sum rule analysis of 
$\gamma$ \cite{ele93}, which is based on the mass splitting in the 
$D$ and $D^*$ isospin dublets, also finds a small value $| \gamma | 
\sim 2.5 \times 10^{-3}$, which is close to the result of Ref. 
\cite{adami1}. Since it is derived from an independent sum rule the 
discrepancy with the result of chiral perturbation theory might have 
a different origin in the $D$ meson channel. This issue and the role 
of instanton corrections in this channel deserve further investigation, 
which will the be subject of a forthcoming publication \cite{for96}. 

\section{Summary and conclusions} 

In this paper we study the role of instantons in the dynamics of 
isospin violation, as it manifests itself in the short distance 
expansion of the nucleon correlation functions. Isospin breaking 
effects lead to differences between the neutron and proton correlators, 
which can be translated via dispersion relations into isopin violating 
vacuum and nucleon parameters.

The isospin-breaking instanton corrections to the 
nucleon correlators show several characteristic qualitative features. 
As a consequence of using Ioffe's interpolating field, instanton 
contributions are absent in the chirally-even amplitudes. Moreover, the 
difference of these amplitudes for neutron and proton is not affected 
by instanton corrections for any choice of the interpolating field. 

The chirally odd amplitude, on the other hand, receives instanton 
contributions of almost the magnitude of the standard OPE terms, as in 
the isosymmetric case. They correct the Wilson coefficients of the unit 
operator and of the quark condensates. The difference between the 
neutron and proton amplitudes is, in fact, mainly generated by the
quark condensate terms, {\it i.e.} by  isopin violating quark modes 
in the vacuum.

This confirms the general expectation that isospin breaking effects 
in hadrons are physically subtle not only because they are small, but 
in particular because they depend sensitively on non-valence-quark 
physics. This is a challenging and little tested regime for hadron 
models, which often neglect vacuum effects alltogether and thus miss 
important sources of isospin asymmetry. Bag (and other quark) model 
calculations which evaluate instanton-induced quark interactions 
between SU(6) states with good isospin fail, for example, to find 
instanton contributions to the neutron-proton mass difference. The 
instanton-corrected OPE, on the other hand, contains vacuum physics 
at short distances and thus provides a more reliable and 
model-independent basis for the study of isospin breaking effects. 

The link between the correlators and nucleon properties 
is established by dispersion relations and takes the form of two QCD 
sum rules for the difference of the neutron and proton amplitudes. In 
adopting an approach for their quantitative analysis one has to decide
between several alternatives. Taking the RHS to be the difference of 
the conventional pole-continuum ans{\"a}tze for the neutron and proton, 
it contains four isospin-breaking parameters which cannot be determined 
independently from a stable fit, even if instanton corrections are taken 
into account. In this situation one can either 
assume relations between these parameters or one can fix the 
only phenomenologically known one, the nucleon mass difference, at 
its experimental value. We adopt the latter approach since it 
does not introduce additional assumptions with uncontrolled 
theoretical errors. 

The resulting sum rules, including the instanton corrections, are 
stable and receive only moderate ($\sim 20 \%$) continuum contributions.
This is a clear improvement over the analogous analysis without 
the instanton terms, where the continuum dominates. 
At the same time, the instanton contributions reduce the difference
between the nucleon pole strengths and enhance the corresponding shift 
in the effective continuum thresholds. Moreover, and perhaps most 
importantly, the optimization of the sum rules with 
direct instanton effects favors larger and more standard values for 
the modulus of the isovector quark condensate, $|\gamma| \simeq 
10^{-2}$, which are close to those found in the chiral analysis.

We also tested an alternative approach towards the sum rule analysis. 
In this case the differences between proton and neutron continuum 
thresholds in both sum rules were assumed to be equal and the 
neutron-proton mass difference $\delta M_N$ was determined from the 
fit. Inclusion of the instanton part allows a consistent fit of both 
sum rules, which seems otherwise impossible. The value of $\delta M_N$ 
is then, however, overestimated by about 80 \%.  This puts the initial 
assumption of an equal deviation of neutron and proton thresholds from 
the isoscalar position into question. It also supports our preference 
for the analysis method discussed above, which does not require ad-hoc 
assumptions to relate fit parameters.

We conclude that instanton corrections play a significant role in the 
analysis of isospin breaking in nucleon sum rules. They, for example,
strongly affect the results for the difference between the nucleon pole 
strengths and for the shifts in the effective continuum thresholds. 
In addition, the instanton corrections enhance the internal consistency 
of the sum rules and predict a larger and more standard value for the 
modulus of the quark condensate difference, $| \gamma |\simeq  (0.8 - 
1) \times 10^{-2}$. 

%%%%%%%%%%%%%%%%%%%%%%%%%%%%%%%%%%%%%%%%%%%%%%%%%%%%%%%%%%%%%%%%%%%%%%
%123456789 123456789 123456789 123456789 123456789 123456789 123456789

%\newpage
\section{Acknowledgements}
H.F. acknowledges support from the U.S. Department of Energy under 
Grant No.\ DE--FG05--93ER--40762 during the initial phase of this work 
at the University of Maryland, and from an European Community Fellowship. 
M.N. acknowledges support from FAPESP and CNPq, Brazil.

%%%%%%%%%%%%%%%%%%%%%%%%%%%%%%%%%%%%%%%%%%%%%%%%%%%%%%%%%%%%%%%%%%%%%%

% \end{thebibliography}
%\eject

\newpage
%%%%%%%%%%%%%%%%%%%%%%%%%%%%%%%%%%%%%%%%%%%%%%%%%%%%%%%%%%%%

\begin{figure}
%\begin{center}
%\epsfysize=\hsize
%\hsize is width of column
%\advance\epsfysize by -0.4cm
%\epsfysize=12.0cm
%\leavevmode
%\setbox\rotbox=\vbox{\epsfbox{npfig1.ps}}\rotl\rotbox
%\end{center}
\caption{Best fit of the RHS (continuous line) of the sum rules to the 
the LHS of the $\Pi_q$ (dotted line) and $\Pi_1$ (dashed line) sum rules.}
\label{fig1}
\end{figure}

\begin{figure}
%\begin{center}
%\epsfysize=\hsize
%\hsize is width of column
%\advance\epsfysize by -0.4cm
%\epsfysize=12.0cm
%\leavevmode
%\setbox\rotbox=\vbox{\epsfbox{c:/papers/n-p-mdif/npfig2.ps}}\rotl\rotbox
%\end{center}
\caption{The neutron-proton mass difference as a function of the Borel 
mass from the optimized $\Pi_q$ (continuous line) and $\Pi_1$ (dashed 
line) sum rules.}
\label{fig2}
\end{figure}

\begin{figure}
%\begin{center}
%\epsfysize=\hsize
%\hsize is width of column
%\advance\epsfysize by -0.4cm
%\epsfysize=12.0cm
%\leavevmode
%\setbox\rotbox=\vbox{\epsfbox{c:/papers/n-p-mdif/npfig3.ps}}\rotl\rotbox
%\end{center}
\caption{Instanton and OPE contributions to the LHS of the $\Pi_1$ 
sum rule. Their sum (dashed line) is fitted to the RHS (continuous 
line).}
\label{fig3}
\end{figure}

\begin{figure}
%\begin{center}
%\epsfysize=\hsize
%\hsize is width of column
%\advance\epsfysize by -0.4cm
%\epsfysize=12.0cm
%\leavevmode
%\setbox\rotbox=\vbox{\epsfbox{c:/papers/n-p-mdif/npfig4.ps}}\rotl\rotbox
%\end{center}
\caption{Same as Fig.1 for the sum rules optimized without instanton
contributions and with $\gamma = - 2 \times 10^{-3}$.}
\label{fig4}
\end{figure}

\begin{figure}
%\begin{center}
%\epsfysize=\hsize
%\hsize is width of column
%\advance\epsfysize by -0.4cm
%\epsfysize=12.0cm
%\leavevmode
%\setbox\rotbox=\vbox{\epsfbox{c:/papers/n-p-mdif/npfig5.ps}}\rotl\rotbox
%\end{center}
\caption{Same as Fig.2 for the sum rules optimized without instanton
contributions and with $\gamma = - 2 \times 10^{-3}$.}
\label{fig5}
\end{figure}
%%%%%%%%%%%%%%%%%%%%%%%%%%%%%%%%%%%%%%%%%%%%%%%%%%%%%%%%%%%%%%%%

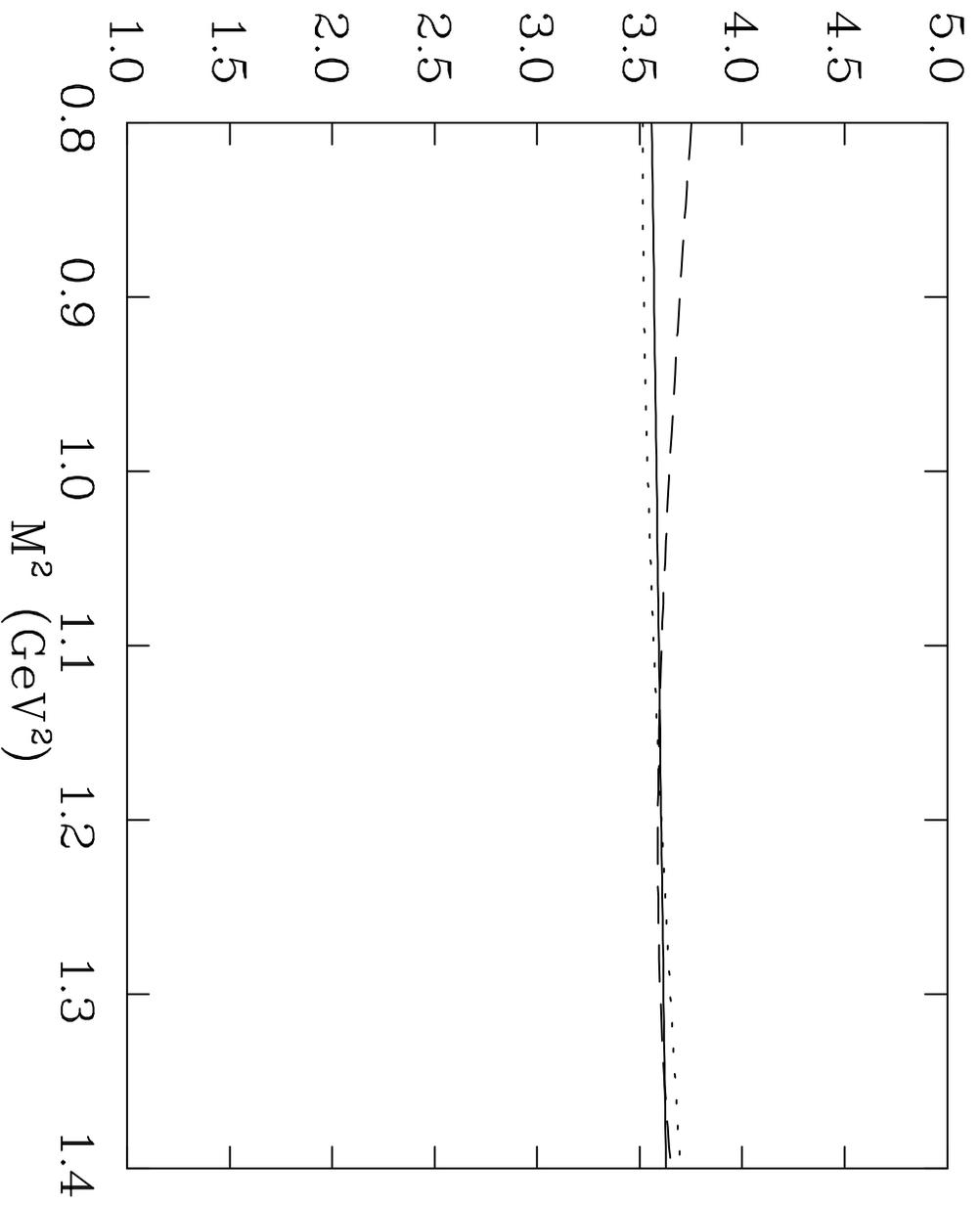
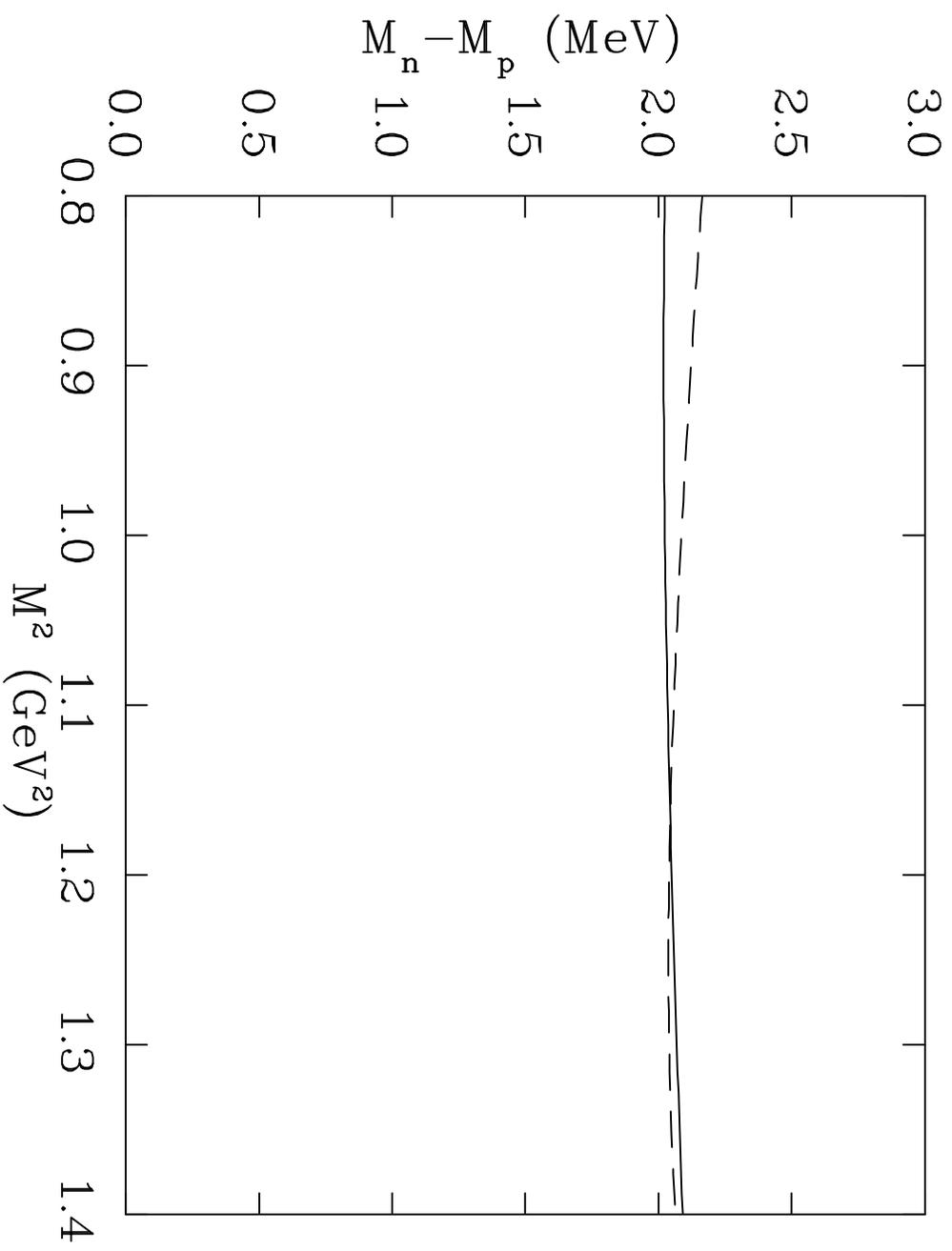
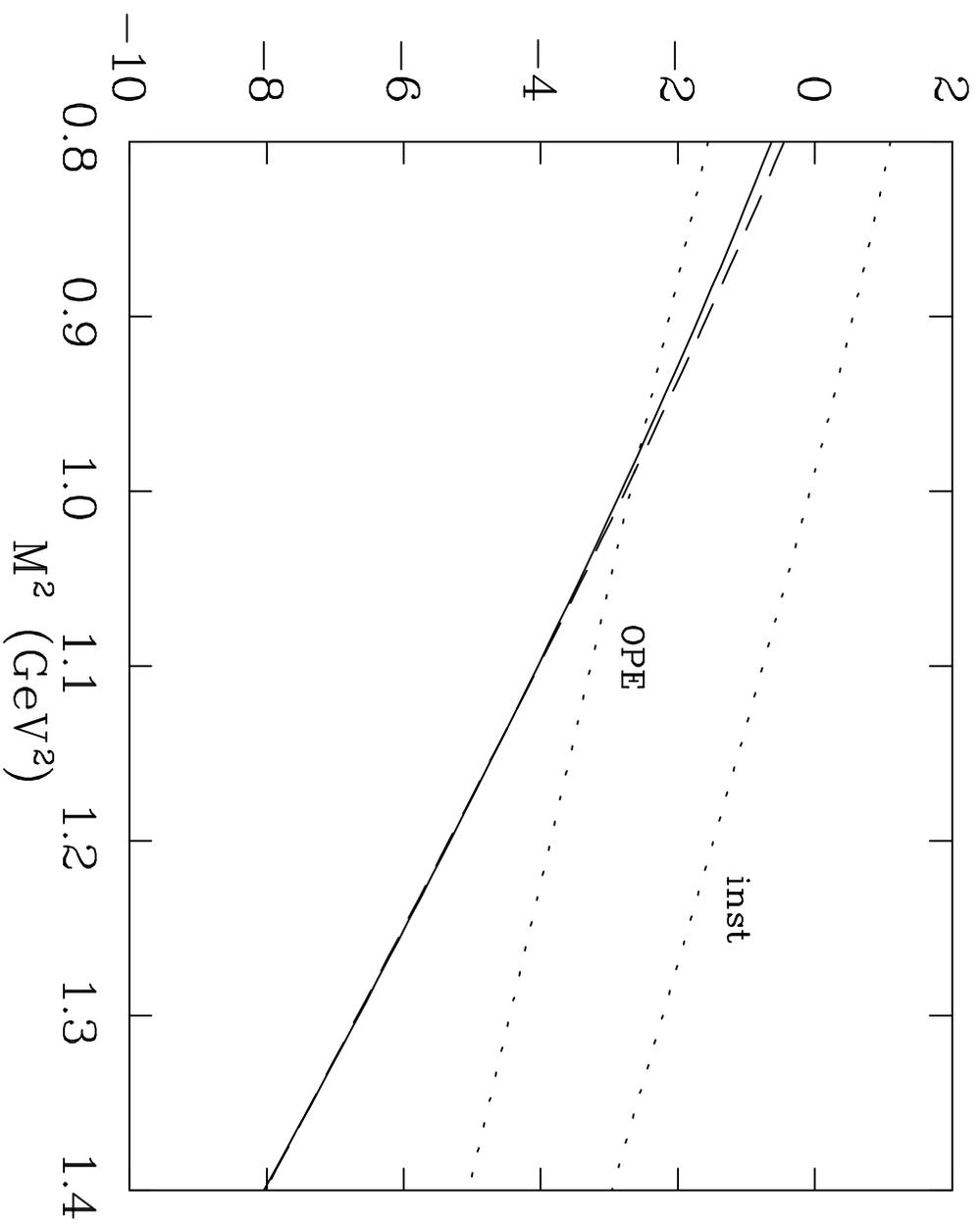
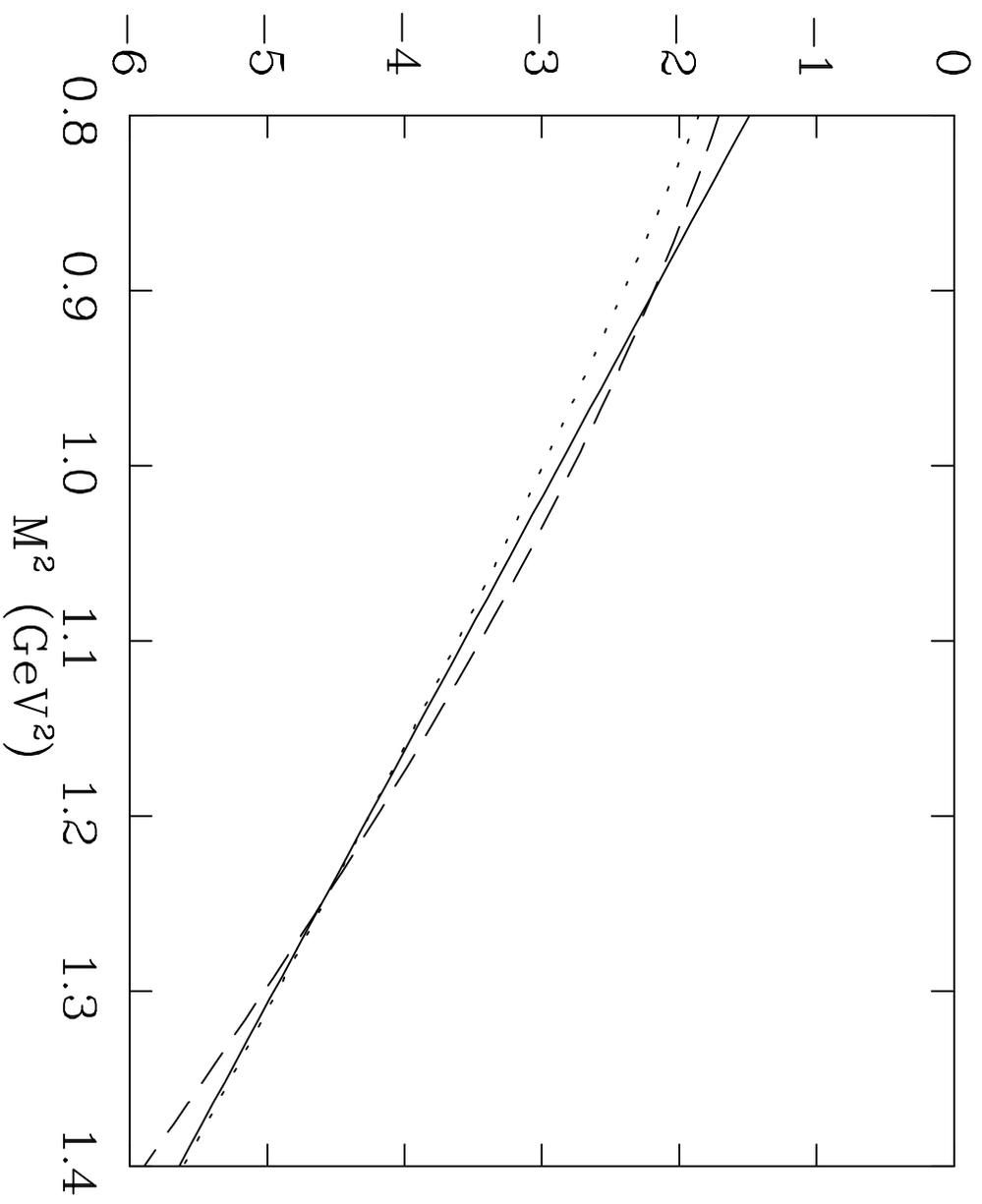
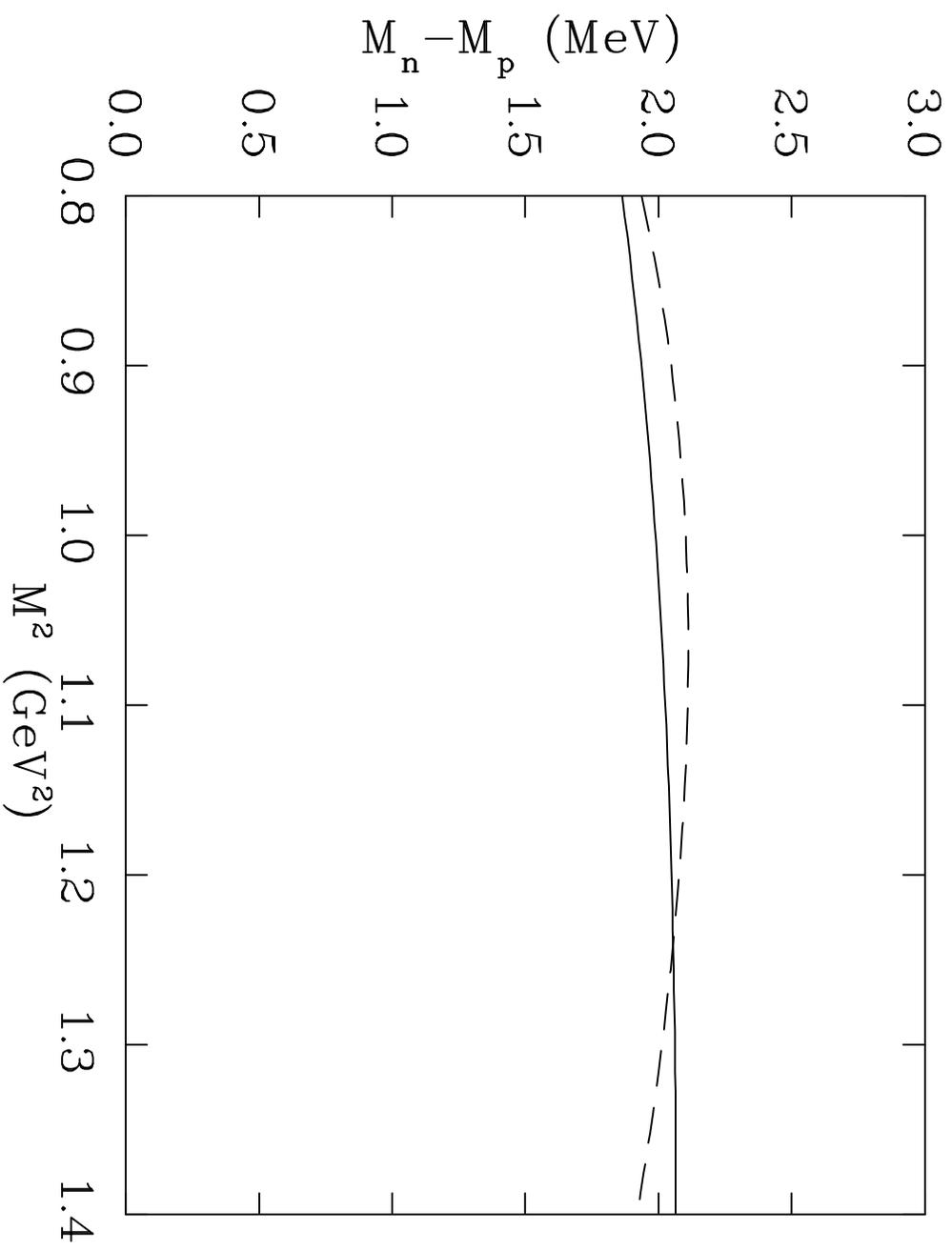
\end{document}